\documentclass[pra,twocolumn,showpacs,preprintnumbers]{revtex4}
\usepackage{amsmath,amssymb,bbm}

\newcommand{\emdash}{---}
\newcommand{\mathd}{\mathrm{d}}
\newcommand{\mathi}{\mathrm{i}}
\newcommand{\mathe}{\mathrm{e}}

\newcommand{\tmem}[1]{{\em #1\/}}

\newcommand{\tmop}[1]{\ensuremath{\text{#1}}}

\begin{document}

\title{Detecting Entanglement in Spatial Interference}

\author{Clemens Gneiting}
\author{Klaus Hornberger}
\affiliation{Max Planck Institute for the Physics of Complex Systems, 
N{\"o}thnitzer Stra{\ss}e 38, 01187 Dresden, Germany}

\date{\today}

\begin{abstract}
We discuss an experimentally amenable class of two-particle states of
motion giving rise to nonlocal spatial interference under position
measurements. Using the concept of modular variables, we derive a
separability criterion which is violated by these non-Gaussian states.
While we focus on the free motion of material particles, the presented
results are valid for any pair of canonically conjugate continuous
variable observables and should apply to a variety of bipartite
interference phenomena.
\end{abstract}

\preprint{\textsf{published in Phys.~Rev.~Lett.~{106}, 210501 (2011)}}
\maketitle

Is it possible to deduce entanglement from an interference pattern, to perform
an ``entangled Young experiment,'' following the famous single-particle
interference experiments? While the wave-particle duality of single material
particles has been a central theme since the early days of quantum mechanics
and is impressively confirmed in interference experiments 
{\cite{ThomsonDavisson1927,Gerlich2007a}}, a similarly convincing
demonstration of quantum nonlocality as implied by entanglement has proven
to be much harder to implement with matter waves.

Although recent experimental progress, in particular in controlling ultracold
atoms, has rendered experiments conceivable that probe entanglement in the free
motion of material particles, a direct implementation of most schemes that have
proven successful with other continuous variable degrees of freedom (e.g., field
modes) fails due to the restricted possibilities to manipulate and detect material
particles. In particular only position measurements are easily doable.
Existing proposals therefore rely either on reduced fluctuations in the center
of mass and relative motion {\cite{ManciniKheruntsyan}}, in the spirit
of Einstein, Podolsky, and Rosen (EPR) \cite{Einstein1935a}, or on the violation
of a Bell inequality {\cite{Bell1964a,Gneiting2008a}}. Both approaches have
drawbacks. The former is based on correlations that appear invitingly easy
to explain in terms of a classical (nonquantum) model, and the latter requires
interferometers to complete the measurements {\cite{Gneiting2008a}}. If one
restricts oneself to elementary position measurements, the states violating
a Bell inequality maximally seem to be hard if not impossible to implement
experimentally {\cite{Kiukas2010a}}.

In view of the great success and the compelling power of single-particle interference
experiments, it is natural to ask whether, instead of violating a Bell inequality,
it is experimentally easier to establish entanglement in the motion of material
particles by means of similarly impressive nonlocal matter wave interference.
We discuss an experimentally amenable class of states which provides such nonlocal
interference. But conceptually, it is not obvious {\tmem{a priori}} that a nonlocal
interference pattern---as intuitively convincing as it may be---can indicate
entanglement, thus strictly excluding the possibility to describe the correlations
in terms of a separable state. While an extensive state tomography could also supply
such a rigorous proof, it would be advantageous to possess an entanglement criterion
in terms of observables that can be directly read off the interference pattern, merely
complemented by measurements of some likewise accessible ``conjugate'' observables.

In this Letter we provide such a criterion. To be more specific, suppose we
hold a two-particle state $\Psi (x_1, x_2)$ which gives rise to a nonlocal
interference pattern when subjected to joint position measurements,
\begin{equation}
  \label{InterferenceEntanglement} | \Psi (x_1, x_2) |^2 = w (x_1 - x_0) w
  (x_2 + x_0) \cos^2 \left( 2 \pi \frac{x_1 - x_2}{\lambda} \right),
\end{equation}
where the envelope $w (x_1 - x_0)$ localizes particle 1 with an uncertainty
$\sigma_x \gg \lambda$ in the vicinity of $x_0$, and similarly particle 2
around $- x_0$. Obviously, the interference pattern describes correlations in
the relative coordinate $x_{\tmop{rel}} = x_1 - x_2$ of the two particles. But
are these correlations necessarily a signature of entanglement? In the case of
EPR states (e.g.,~squeezed Gaussian states), entanglement can be deduced from
the reduced fluctuations in both the relative coordinate $x_{\tmop{rel}}$ and
the total momentum $p_{\tmop{tot}} = p_1 + p_2$, since the canonically
conjugate operator pairs $\mathsf{x}_j$, $\mathsf{p}_j$ ($[ \mathsf{x}_j,
\mathsf{p}_j] = \mathi \hbar$), $j = 1, 2$, set lower limits to these
fluctuations for separable states {\cite{Duan2000a,Simon2000a}}. In the
situation described by (\ref{InterferenceEntanglement}), in contrast, it is
not the relative coordinate that is ``squeezed,'' but its value modulo
$\lambda$.

We show how this observation can be employed to derive an
entanglement criterion. The key is to identify {\tmem{modular variables}}
{\cite{Aharonov1969a}} as the appropriate pair of conjugate observables.
The criterion is rooted in a state-independent additive uncertainty relation
(UR) for these variables, which remedies the problems arising from the
operator-valued commutator appearing in the Robertson UR. We construct a class
of non-Gaussian states, denoted {\tmem{modular entangled states}}, which offer
natural and robust generation protocols and violate this criterion. The
interference pattern in (\ref{InterferenceEntanglement}) is shown to represent
only the weakest form of nonlocal correlations exhibited by this class.

{\tmem{Multislit interference.}}{\emdash}To discuss the prerequisites of
particle interference and its relation to the modular variables it is
instructive to recapitulate the single-particle case first. Ideally, the
(transverse) state immediately after passing an aperture of $N$ slits is
described by a superposition of $N$ spatially distinct state components
determined by the shape of the slits,
\begin{equation}
  \label{GridState} \langle x| \psi_{\tmop{MS}} \rangle = \frac{1}{\sqrt{N}} 
  \sum^{N - 1}_{n = 0} \langle x + n L| \psi \rangle,
\end{equation}
where $L$ denotes the slit separation. The particular shape of the single-slit
wave function $| \psi \rangle$ is irrelevant for our discussion provided its
spatial width $\sigma_x$ satisfies $\sigma_x \ll L$. This guarantees that the
envelope of the resulting fringe pattern varies slowly on the scale of a
single fringe period and thus encloses a large number of interference fringes.
Note that the state (\ref{GridState}) can equally be read as a longitudinal
superposition of comoving wave packets.

The subsequent dispersive spreading of the $N$ wave packets during the free
propagation to the screen results in their overlap and interference, yielding
in the far-field limit the characteristic interference pattern on the screen.
In terms of the initial state (\ref{GridState}), this position measurement at
asymptotic times corresponds to a formal momentum measurement of $p = m (x -
\langle \psi_{\tmop{MS}} | \mathsf{x} | \psi_{\tmop{MS}} \rangle) / t$. The
probability distribution
\begin{equation}
  \label{GridFringePattern} | \langle p| \psi_{\tmop{MS}} \rangle |^2 = |
  \langle p| \psi \rangle |^2 F_N \left( \frac{p L}{h} \right)
\end{equation}
exhibits the fringe pattern $F_N (x) = 1 + (2 / N) \sum^{N - 1}_{j = 1} (N -
j) \cos \left( 2 \pi j x \right)$. In case of $N = 2$ Eq.
(\ref{GridFringePattern}) reduces to the sinusoidal fringe pattern of the
double slit, whereas for $N > 2$ one obtains the sharpened main maxima and
suppressed side maxima characteristic for multislit interference. This
reflects a tradeoff between the number of superposed wavepackets $N$ and the
uncertainty of the phase of the interference pattern, in analogy to the
tradeoff between the variances of a conjugate variable pair. A similar
tradeoff exists between the number of fringes $M \approx \sigma_p L / h \approx L
/ \sigma_x$ covered by the envelope of the interference pattern and the
width-to-spacing ratio $\sigma_x / L \approx 1 / M$.

{\tmem{Modular variables.}}{\emdash}These mutual relationships between the
multislit state (\ref{GridState}) and the resulting interference pattern
(\ref{GridFringePattern}) are captured best by splitting the position
(momentum) operator into an integer component $\mathsf{N}_x$ ($\mathsf{N}_p$)
and a modular component $\overline{\mathsf{x}}$ ($\overline{\mathsf{p}}$)
{\cite{Aharonov1969a}},
\begin{eqnarray}
  \mathsf{x} = \mathsf{N}_x \ell + \overline{\mathsf{x}}, & & \mathsf{p} =
  \mathsf{N}_p  \frac{h}{\ell} + \overline{\mathsf{p}}, 
  \label{ModularVariables}
\end{eqnarray}
where $\bar{x} = (x + \ell / 2) \tmop{mod} \ell - \ell / 2$ and $\bar{p} = (p
+ h / 2 \ell) \tmop{mod} (h / \ell) - h / 2 \ell$. (For convenience, we define
the modular variables symmetrically with respect to the origin.) Recent
applications of the modular variables are discussed in
{\cite{Tollaksen2010a,Popescu2010a,Plastino2010a}}.

For the multislit state (\ref{GridState}) the adequate choice of the
partition scale is given by $\ell = L$. The probability distribution
(\ref{GridFringePattern}) can then be written as $| \langle p = N_p h / L +
\bar{p} | \psi_{\tmop{MS}} \rangle |^2 \approx | \langle p = N_p h / L| \psi
\rangle |^2 F_N ( \bar{p} L / h)$, which indicates that the modular variables
isolate different characteristic aspects of interference: the periodic fringe
pattern is described by the modular momentum $\bar{p}$, its envelope by the
integer momentum $N_p$. Similarly, $N_x$ describes the distribution of wave
packets in (\ref{GridState}) and $\bar{x}$ their (common) shape.

The modular variables $\overline{\mathsf{x}}$, $\overline{\mathsf{p}}$ have
the remarkable property that they commute, $[ \overline{\mathsf{x}},
\overline{\mathsf{p}}] = 0$, despite originating from conjugate observables
{\cite{Aharonov1969a,Busch1987a}}. The common eigenstates $| \bar{x}, \bar{p}
\rangle$ of $\overline{\mathsf{x}}$ and $\overline{\mathsf{p}}$ with
eigenvalues $\bar{x}$ and $\bar{p}$ read $| \bar{x}, \bar{p} \rangle =
\sqrt{\ell / h}  \sum_{n \in \mathbbm{Z}} \mathe \tmop{xp} (\mathi \bar{p} n
\ell / \hbar) |n \ell + \bar{x} \rangle_x$, or, equivalently, $| \bar{x},
\bar{p} \rangle = \sqrt{1 / \ell} \mathe \tmop{xp} (- \mathi \mathsf{p} 
\bar{x} / \hbar) \sum_{m \in \mathbbm{Z}} |m h / \ell + \bar{p} \rangle_p$.
The tradeoff between the number of superposed wave packets $N$ and the phase
of the interference pattern is now reflected by a conjugate relationship
between the integer position $\mathsf{N}_x$ and the modular momentum
$\overline{\mathsf{p}}$,
\begin{equation}
  \label{ModularMomentumCommutator} [ \mathsf{N}_x, \overline{\mathsf{p}}] =
  \frac{\mathi \hbar}{\ell}  \left( \mathbbm{l}- \frac{h}{\ell} \int^{\ell /
  2}_{- \ell / 2} \mathd \bar{x} | \bar{x}, \bar{p} = h / 2 \ell \rangle
  \langle \bar{x}, \bar{p} = h / 2 \ell | \right) .
\end{equation}
Similarly, the tradeoff between the width-to-spacing ratio $\sigma_x / L$ and
the number of covered fringes is described by the commutator of the modular
position $\overline{\mathsf{x}}$ and the integer momentum $\mathsf{N}_p$,
\begin{equation}
  \label{ModularPositionCommutator} [ \overline{\mathsf{x}}, \mathsf{N}_p] =
  \frac{\mathi \ell}{2 \pi} \left( \mathbbm{l}- \ell \int^{h / 2 \ell}_{- h /
  2 \ell} \mathd \bar{p} | \bar{x} = \ell / 2, \bar{p} \rangle \langle \bar{x}
  = \ell / 2, \bar{p} | \right) .
\end{equation}
The projection operators on the right-hand side of
(\ref{ModularMomentumCommutator}) and (\ref{ModularPositionCommutator}) result
from the boundedness of the modular variables and are indispensible to ensure
the validity of the Robertson UR. This is similar to the relationship between
an angular position operator and its conjugate angular momentum
{\cite{Franke-Arnold2004a}}.

{\tmem{Squeezed modular position states.}}{\emdash}The multislit states
(\ref{GridState}) display their interference in momentum. In view of the
symmetry between the two pairs $( \mathsf{N}_x, \overline{\mathsf{p}})$ and $(
\overline{\mathsf{x}}, \mathsf{N}_p)$, one can construct another class of
states where the modular variables exchange their roles. This is achieved by
superposing wave packets that are distinct in momentum (instead of
position),
\begin{equation}
  \label{ModularPositionSqueezedStates} | \psi_{\tmop{SMP}} \rangle =
  \frac{1}{\sqrt{N}}  \sum^{N - 1}_{n = 0} | \psi_{x_0, (N_0 + n) h / \lambda}
  \rangle,
\end{equation}
where $\langle x| \psi_{x_0, p_0} \rangle = \phi \left( x - x_0 \right) \mathe
\tmop{xp} [\mathi p_0 (x - \bar{x}_0) / \hbar]$ denotes a (well-behaved) wave
packet that is localized in phase space around $(x_0, p_0)$. $N_0$ represents
an arbitrary base integer momentum. Distinctness of the wave packets requires
that their momentum width $\sigma_p$ is smaller than their separation in
momentum space, $\sigma_p \ll h / \lambda$, or, equivalently, $\sigma_x \gg
\lambda$. A hypothetical ``momentum grid'' with slit width $h / d$, $d \gg
\lambda$, could, e.g., prepare a state with $\phi (x) = \tmop{sinc} \left[ \pi
x / d \right] / \sqrt{d}$.

A position measurement of the states (\ref{ModularPositionSqueezedStates})
reveals an interference pattern with periodicity $\lambda$, $| \langle x|
\psi_{\tmop{SMP}} \rangle |^2 = | \phi \left( x - x_0 \right) |^2 F_N ((x -
\bar{x}_0) / \lambda)$. Note that $\bar{x}_0$ determines the phase of the
fringe pattern. Increasing $N$ results in the formation of sharp main maxima,
which is reflected by the decreasing variance of the modular position variable
(now with $\ell = \lambda$),
\begin{equation}
  \label{ModularPositionVariance} \langle (\Delta \overline{\mathsf{x}})^2
  \rangle_{\tmop{SMP}} = \frac{\lambda^2}{12} [1 - S_1 (N)] .
\end{equation}
In this sense one may denote the states (\ref{ModularPositionSqueezedStates})
as {\tmem{squeezed modular position}} {\tmem{states}}. The monotonically
increasing squeezing function $S_1 (N) = - (12 / \pi^2) \sum_{j = 1}^{N - 1}
(- 1)^j (N - j) / N j^2 < 1$ (for $\bar{x}_0 = 0$) is evaluated in Table
\ref{SqueezingFunctionEvaluation} for representative $N$. Notably, in the
limit $N \rightarrow \infty$ the variance (\ref{ModularPositionVariance})
vanishes, indicating perfect squeezing.\begin{table}[tb]
  \begin{tabular}{l|llllll}
    $N$ & 1 & 2 & 3 & 4 & 10 & 100\\ \hline
    $S_1 (N)$ & 0.0{\phantom{0}} & 0.61 & 0.71 & 0.79 & 0.92 & 0.99\\
    $S_2 (N)$ & 0.0 & 0.30 & 0.46 & 0.55 & 0.76 & 0.96
  \end{tabular}
  \caption{\label{SqueezingFunctionEvaluation}Evaluation of the squeezing
  functions $S_1 (N)$ and $S_2 (N)$ for several superposition ranks $N$. $S_1
  (N)$ and $S_2 (N)$ describe the squeezing of the modular position $\bar{x}$
  in the single-particle case (\ref{ModularPositionVariance}) and of the
  modular relative position $\bar{x}_1 - \bar{x}_2$ in the two-particle
  entangled case (\ref{ModularRelativePositionVariance}), respectively.}
\end{table}

Correspondingly, the variance of the integer momentum $\mathsf{N}_p$ increases
with $N$, $\langle (\Delta \mathsf{N}_p)^2 \rangle_{\tmop{SMP}} = (N^2 - 1) /
12$. For $N = 1$, however, $\langle (\Delta \mathsf{N}_p)^2 \rangle_{\tmop{SMP}}$
vanishes (since the $| \psi_{x_0, p_0} \rangle$ are localized on the scale of
the integer momentum), while $\langle (\Delta \overline{\mathsf{x}})^2
\rangle_{\tmop{SMP}}$ remains finite according to (\ref{ModularPositionVariance}).
Validity of the Robertson UR, $\langle (\Delta \mathsf{N}_p)^2 \rangle_{\tmop{SMP}}
\langle (\Delta \overline{\mathsf{x}})^2 \rangle_{\tmop{SMP}} \geqslant |
\langle [ \overline{\mathsf{x}}, \mathsf{N}_p] \rangle_{\tmop{SMP}} |^2 / 4$,
thus requires that the projector on the right-hand side of
(\ref{ModularPositionCommutator}) renders the Robertson UR trivial for $N =
1$. Indeed, we find $| \langle [ \overline{\mathsf{x}}, \mathsf{N}_p]
\rangle_{\tmop{SMP}} | = (\ell/2 \pi)  \left[ 1 - (1 + (- 1)^{N + 1})/2 N \right],$
which vanishes for $N = 1$. This irrelevance of the Robertson UR in the case
$N = 1$ impedes its employment in the separability criterion presented below.
Note that $| \langle [ \overline{\mathsf{x}}, \mathsf{N}_p] \rangle_{\tmop{SMP}} |$
converges towards the canonical constant value, while the projector term in
(\ref{ModularPositionCommutator}) is still relevant for $N = 3$. Its
alternating structure can be traced back to either minima or (side) maxima of
the fringe pattern coinciding with $\bar{x} = \lambda / 2$.

As an advantage of the modular position squeezed states
(\ref{ModularPositionSqueezedStates}) compared to the modular momentum
squeezed states (\ref{GridState}), they exhibit interference by immediate
position measurements, while to determine the integer momentum one must only
distinguish the macroscopically distinct components $| \psi_{x_0, p_0}
\rangle$, which is easy once they are sufficiently separated by free
propagation. At the same time, as superpositions of different velocities, they
are genuine matter wave states without photonic analogue.

\enlargethispage{\baselineskip}

{\tmem{Modular entangled states.}}{\emdash}We are now prepared to move on to
entangled states of two material particles. Ultimately, we are interested in
states that reveal their entanglement by a nonlocal interference pattern
similar to (\ref{InterferenceEntanglement}). To this end, we introduce
two-particle {\tmem{modular position entangled}} (MPE) {\tmem{states}}, which
are defined by superposing correlated pairs of (counterpropagating) wave
packets of different velocities,
\begin{equation}
  \label{ModularPositionEntangledStates} | \Psi_{\tmop{MPE}} \rangle =
  \frac{1}{\sqrt{N}}  \sum^{N - 1}_{n = 0} | \psi_{x_0, (N_0 + n) h / \lambda}
  \rangle_1 | \psi_{- x_0, - (N_0 + n) h / \lambda} \rangle_2 .
\end{equation}
Only for clarity we assume that the particles are spatially separated,
positioned at $\pm x_0$. Moreover, it is clear that one could equally define
modular momentum entangled (MME) states. Such states could be generated (to
good approximation) by the sequential coherent dissociation of a diatomic
molecule {\cite{Gneiting2010b}}. For convenience we consider a superposition
of product states; correlated components $| \Psi_{x_0, p_0 ; - x_0, - p_0}
\rangle$ would not modify our conclusions, since the latter are based on
entangled integer momenta $\mathsf{N}_{p, j}$, i.e.,~distinctive ``bulk''
properties of the particles. This is in contrast to EPR states, where the
relevant correlations reside in the microscopic fluctuations of the center of
mass and relative variables. Performing position measurements on each side,
the nonseparable structure of (\ref{ModularPositionEntangledStates}) gives
rise to an interference pattern in the relative position $x_{\tmop{rel}}$,
$| \langle x_1, x_2 | \Psi_{\tmop{MPE}} \rangle |^2 = | \phi \left( x_1 - x_0
\right) |^2 | \phi \left( x_2 + x_0 \right) |^2 F_N ((x_1 - x_2) / \lambda)$
(with $\bar{x}_{0, 1} = \bar{x}_{0, 2}$), or, equivalently, to a
squeezing in the modular relative position $\bar{x}_{\tmop{rel}} = \bar{x}_1 -
\bar{x}_2$.

{\tmem{Modular entanglement criterion.}}{\emdash}The correlations in
$\bar{x}_{\tmop{rel}}$ and the total integer momentum $N_{p, \tmop{tot}}
= N_{p, 1} + N_{p, 2}$ exhibited by the MPE states
(\ref{ModularPositionEntangledStates}) can be exploited to demonstrate the
underlying entanglement. In analogy to {\cite{Duan2000a}}, we consider the sum
of variances, $\langle (\Delta \mathsf{N}_{p, \tmop{tot}})^2 \rangle_{\rho} +
\langle (\Delta \overline{\mathsf{x}}_{\tmop{rel}})^2 \rangle_{\rho} /
\ell^2$. Using the Cauchy-Schwarz inequality one can show that a separable
state of motion, $\rho = \sum_i p_i \rho_{1 i} \otimes \rho_{2 i}$, implies
$\langle (\Delta \mathsf{N}_{p, \tmop{tot}})^2 \rangle_{\rho} + \langle
(\Delta \overline{\mathsf{x}}_{\tmop{rel}})^2 \rangle_{\rho} / \ell^2
\geqslant \sum_{i, j} p_i \{\langle (\Delta \mathsf{N}_{p, j})^2 \rangle_i +
\langle (\Delta \overline{\mathsf{x}}_j)^2 \rangle_i / \ell^2 \}$, with $j =
1, 2$. In contrast to {\cite{Duan2000a}}, we cannot use the Robertson UR to
estimate the remaining sums of variances, since the expectation value of the
state-dependent commutator (\ref{ModularPositionCommutator}) vanishes when
evaluated for an MPE state (\ref{ModularPositionEntangledStates}). However,
one can establish a state-independent additive uncertainty relation
for the modular variables, $\langle (\Delta \mathsf{N}_{p, j})^2 \rangle +
\langle (\Delta \overline{\mathsf{x}}_j)^2 \rangle / \ell^2 \geqslant
c_{\mathsf{N}_p, \overline{\mathsf{x}}} > 0$. Using this, we immediately get
the desired criterion,
\begin{equation}
  \label{ModularEntanglementCriterion} \langle (\Delta \mathsf{N}_{p,
  \tmop{tot}})^2 \rangle_{\rho} + \frac{1}{\ell^2} \langle (\Delta
  \overline{\mathsf{x}}_{\tmop{rel}})^2 \rangle_{\rho} \geqslant 2
  c_{\mathsf{N}_p, \overline{\mathsf{x}}},
\end{equation}
which must be satisfied by any separable state. The constant $c_{\mathsf{N}_p,
\overline{\mathsf{x}}}$ is given by the smallest eigenvalue $\mu_0$ of the
operator $\mathsf{A}_j = \mathsf{N}_{p, j}^2 + \overline{\mathsf{x}}^2_j /
\ell^2$. The corresponding differential equation in the common eigenbasis of
$\overline{\mathsf{x}}_j$ and $\overline{\mathsf{p}}_j$ is solved by $\psi (
\bar{x}_j, \bar{p}_j) = \exp (- \pi \bar{x}_j^2 / \ell^2) M (- \pi \mu / 2 + 1
/ 4, 1 / 2, 2 \pi \bar{x}_j^2 / \ell^2) \chi ( \bar{p}_j)$, with $M (a, b ;
x)$ the Kummer function and $\chi ( \bar{p})$ arbitrary. Continuity requires
a vanishing first derivative at $\bar{x}_j = \ell / 2$, which implicitly
determines the (discrete) spectrum $\{\mu_{\nu} \}$ of $\mathsf{A}_j$.
Its smallest eigenvalue evaluates numerically as $c_{\mathsf{N}_p,
\overline{\mathsf{x}}} \cong 0.078235$; second order perturbation theory
(with $\mathsf{A}_j$ expressed in the common eigenbasis of $\mathsf{N}_{p, j}$,
$\overline{\mathsf{p}}_j$) yields a reasonable analytic approximation,
$c_{\mathsf{N}_p, \overline{\mathsf{x}}} \approx 1 / 12 (1 - 1 / 15)$.
We note that a criterion similar to (\ref{ModularEntanglementCriterion})
can be established for $\mathsf{N}_{x, \tmop{tot}}$ and
$\overline{\mathsf{p}}_{\tmop{rel}}$.

The MPE states (\ref{ModularPositionEntangledStates}) violate the separability
criterion (\ref{ModularEntanglementCriterion}) for any $N \geqslant 2$.
Indeed, the resulting variances read $\langle (\Delta \mathsf{N}_{p,
\tmop{tot}})^2 \rangle_{\tmop{MPE}} = 0$ and
\begin{equation}
  \label{ModularRelativePositionVariance} \langle (\Delta
  \overline{\mathsf{x}}_{\tmop{rel}})^2 \rangle_{\tmop{MPE}} =
  \frac{\lambda^2}{6} [1 - S_2 (N)],
\end{equation}
(again with $\ell = \lambda$) where the monotonically increasing squeezing
function $S_2 (N) = (6 / \pi^2) \sum_{j = 1}^{N - 1} (N - j) / N j^2 < 1$ is
evaluated in Table \ref{SqueezingFunctionEvaluation} for several
representative $N$. This proves the possibility to deduce entanglement from a
nonlocal interference pattern. Again, one can achieve perfect squeezing in the
limit $N \rightarrow \infty$; the interference pattern
(\ref{InterferenceEntanglement}) corresponds to $N = 2$.

The MPE states (\ref{ModularPositionEntangledStates}) (and MME states alike)
generalize single-particle interferometric schemes such as double-slit or
grid experiments to the case of two entangled particles. Aside from the
additional requirement to provide the correlations between the particles,
the MPE states thus inherit both the advantages and the challenges of such
schemes. Similar to any interference experiment, the phase $\bar{x}_0$
of the superposed components $| \psi_{x_0, p_0} \rangle$ must be well controlled,
and also all components should share the same shape $\phi (x)$ [see
(\ref{ModularPositionSqueezedStates})]. (On the other hand, the particular shape
is to a large extent irrelevant, which leaves it to the experimenter to choose
easily producible states.) Deviations from these conditions result
in a visibility-reducing blurring of the fringe pattern and thus
in an attenuation of the squeezing of the modular variable. However, a simple
robustness check, where the MPE states are mixed with merely classically
(integer momentum) correlated states, reveals that for $N = 2$ a classically
correlated admixture of up to $79\%$ would sustain the violation of the
separability criterion, corresponding to a fringe visibility of merely $21\%$.
This robustness, which even improves with increasing $N$, underlines the
appropriateness of the separability criterion (\ref{ModularEntanglementCriterion})
to capture entanglement in spatial interference, and it should leave sufficient
freedom to cope with possible experimental limitations.

A realistic generation protocol for MPE states would, e.g., gradually dissociate
an ultracold diatomic Feshbach molecule such that subsequent dissociation instants
produce wave packets with staggered kinetic energies {\cite{Gneiting2010b}}.
Appropriate dissociation pulses can achieve that all of these consecutive wave
packets meet simultaneously on each side. This constitutes an approximate MPE state,
where the superposed $| \psi_{x_0, p_0} \rangle$ then realistically differ by
different stages of dispersion. We checked for $N = 2$ and lithium atoms that
this dispersion-induced shape difference can easily be kept under control with
realistic parameters, yielding an experimentally resolvable fringe pattern with
$\lambda \approx 100 \mu \text{m}$ and a visibility of $85\%$. On the other hand,
a ``grid state preparation'' of transversal MME states, starting, e.g., with an
EPR correlated particle pair and then each particle passing a grating, would
provide the identity in phase and shape of the $| \psi_{x_0, p_0} \rangle$
for any $N$ by means of the state preparation.

{\tmem{Conclusion.}}{\emdash}We presented a scheme to provide and detect
entanglement in the motion of two free material particles. Elementary position
measurements at macroscopically distinct sites give rise to a nonlocal interference
pattern; the nonseparability then follows from reduced fluctuations in
adapted modular variables. In this sense, the scheme allows one to ``deduce
entanglement from interference'', and hence to illustrate the wave-particle
duality on a new level including quantum mechanical nonlocality.

We emphasize that the modular variables are merely a matter of interpretation
in our scheme and can be deduced from ordinary position and momentum measurements.
Finally, it is clear that the entanglement criterion is applicable to any bipartite
continuous variable system with conjugate operator pairs, e.g.,~quadrature amplitudes
of field modes, and could thus offer a valuable alternative to existing entanglement
detection schemes. Homodyning entangled coherent states {\cite{Sanders1992a}} may
serve as an immediate example.


\end{document}